\documentclass[fleqn,10pt]{wlscirep}
\usepackage[utf8]{inputenc}
\usepackage[T1]{fontenc}
\title{Connecting complex networks to nonadditive entropies}

\author[1]{R. M. de Oliveira}
\author[2,*]{Samuraí Brito}
\author[1,3]{L. R. da Silva}
\author[3,4,5,6]{Constantino Tsallis}
\affil[1]{Federal University of Rio Grande do Norte, Departamento de F\'isica Te\'orica e Experimental, Natal-RN, 59078-900, Brazil.}
\affil[2]{International Institute of Physics, Federal University of Rio Grande do Norte, 59070-405 Natal-RN, Brazil.}
\affil[3]{National Institute of Science and Technology of Complex Systems, Brazil.}

\affil[4]{Centro Brasileiro de Pesquisas F\'isicas, Rua Xavier Sigaud 150, 22290-180 Rio de Janeiro-RJ, Brazil.}

\affil[5]{Santa Fe Institute, 1399 Hyde Park Road, New Mexico 87501, USA.}

\affil[6]{Complexity Science Hub Vienna, Josefstaedter Strasse 39, A 1080 Vienna, Austria.}

\affil[*]{samuraigab@gmail.com}

\begin{abstract}
Boltzmann-Gibbs statistical mechanics applies satisfactorily to a plethora of systems. It fails however for complex systems generically involving strong space-time entanglement. Its generalization based on nonadditive $q$-entropies adequately handles a wide class of such systems. We show here that scale-invariant networks belong to this class. We numerically study a $d$-dimensional geographically located network with weighted links and exhibit its 'energy' distribution per site at its quasi-stationary state. Our results strongly suggest 
a correspondence between the random geometric problem and a class of thermal problems within the generalised thermostatistics. The Boltzmann-Gibbs exponential factor is generically substituted by its $q$-generalisation, and is recovered in the $q=1$ limit when the nonlocal effects fade away. The present connection should cross-fertilise experiments in both research areas.
\end{abstract}
\begin{document}

\flushbottom
\maketitle
%
%
\thispagestyle{empty}


\section*{Introduction}

Boltzmann-Gibbs (BG) statistical mechanics constitutes one of the pillars of contemporary theoretical physics. As such is has uncountable successes for a great variety of physical systems. However, when the system constituents have a generically strong space-time entanglement, this theory does not apply. Such is the case already pointed in 1902 by Gibbs himself, namely when the standard partition function diverges, e.g., gravitation.
It is in this context that it was proposed in 1988 ~\cite{Tsallis1988} the generalisation -- hereafter referred to as {\it nonextensive statistical mechanics} -- of the BG theory based on nonadditive entropies, namely $S_q = k \frac{1-\sum_i p_i^q}{q-1}$, which recovers $S_{BG}=-k \sum_{i} p_i \ln p_i$ in the $q\to 1$ limit. The composition of two probabilistically independent systems $A$ and $B$ yields straightforwardly $S_q(A+B)/k= [S_q(A)/k] + [S_q(B)/k] +(1-q) [S_q(A)/k][ S_q(B)/k] $. As we see, the BG entropic additivity is recovered when $q=1$. The fundamental advantage associated with $q \ne 1$ is that, for strongly correlated systems, it enables, as illustrated in  ~\cite{TsallisGellMannSato2005}, the preservation of the {\it extensivity} of the thermodynamic entropy, mandated in all circumstances by the Legendre structure of classical thermodynamics.

In parallel with the above, the study of complex networks has been intensified around the world ~\cite{Price1965,WattsStrogatz1998,BarabasiAlbert1999,AlbertBarabasi2002,Newman2003}. 
Networks can be found everywhere. Society is formed by humans linked through relationships. The Internet is a set of devices communicating with each other. The brain is formed by neurones communicating through synapses. 
All these completely different systems can be translated onto a simple set of \emph{nodes} (or \emph{sites}) and \emph{edges} (or \emph{links}) obeying some connection rule, and the tools of network science can be successfully used to study them.
Typical applications of this area can be found in classical and quantum internet ~\cite{tilch2020multilayer,BritoQN2020}, medicine ~\cite{Goh8685,Gomes2014}, neuroscience~\cite{Mota2017}, and sociology ~\cite{Nagler2011,Shirado2017}. It was thought, during more than a decade, that most of real networks were purely scale-free meaning that the distribution of the number of links in the network follows a power-law distribution. It was recently argued that most real networks are not pure scale-free~\cite{Broido2019}, paving the way for new possibilities to describe them.

During the initial years, network science and nonextensive statistical mechanics were seen as completely different areas. But meaningful connections started in 2005 ~\cite{SoaresTsallisMarizSilva2005, ThurnerTsallis2005, Thurner2005, BritoSilvaTsallis2016, NunesBritoSilvaTsallis2017, BritoNunesSilvaTsallis2019, CinardiRapisardaTsallis2020}.
It is nowadays known that the degree distribution of asymptotically scale-free networks at the thermodynamic limit is of the form $P(k) \propto e_q^{-k/\kappa}$, where the $q$-exponential function is defined as $e^{z}_q\equiv [1 + (1-q)z]^{\frac{1}{1-q}}$ ($e_1^z=e^z$). This form, more precisely $p_q(\varepsilon_i) = e_q^{-\beta_q\,\varepsilon_i}/Z_q$,
optimizes the entropy $S_q$ under appropriate canonical constraints, $\varepsilon_i$ being the site energy and $\beta_q$ the inverse temperature; the BG weight is recovered at the $q\to 1$ limit. This thermostatistical approach has been successfully applied in a wide diversity of areas, such as long-range-interacting Hamiltonian systems \cite{CirtoRodriguezNobreTsallis2018}, vortices in type II supercondutors \cite{AndradeSilvaMoreiraNobreCurado2010}, cold atoms \cite{LutzRenzoni2013}, granular matter \cite{CombeRichefeuStasiakAtman2015}, high-energy physics experiments on Earth \cite{WongWilk2013} and observations in the outer space \cite{YalcinBeck2018,SmollaSchaferLeschBeck2020}, civil engineering \cite{GrecoTsallisRapisardaPluchinoFicheraContrafatto2020}, and for predicting COVID-19 peaks around the world \cite{TsallisTirnakli2020,TirnakliTsallis2020}.

\begin{figure}
\centering
\includegraphics[width=\textwidth]{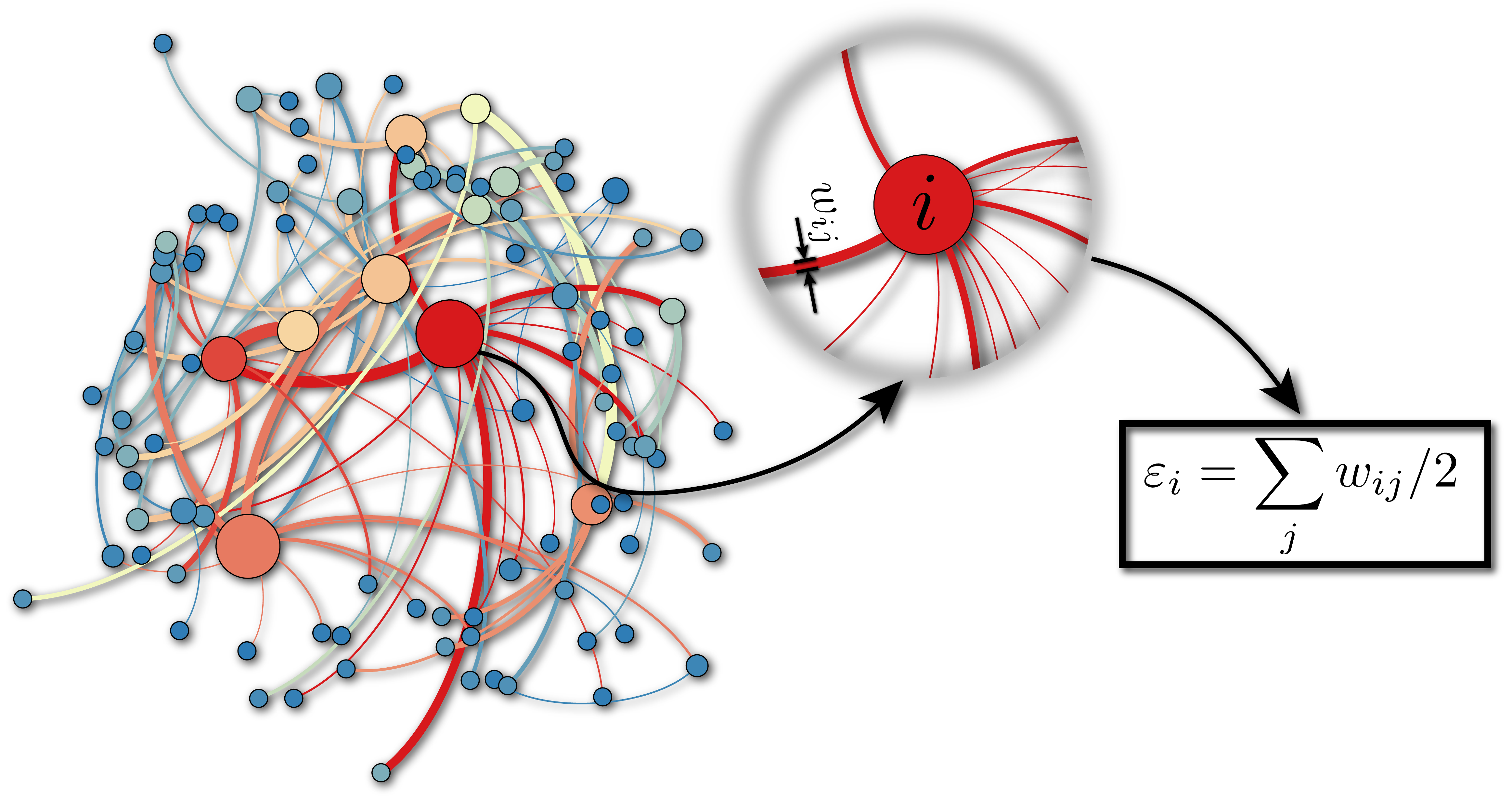}
\caption{Sample of a $N=100$ network for $(d,\alpha_A,\alpha_G,\eta,w_0) = (2,1,5,1,1)$. As can be seen, for this choice of parameters, hubs (highly connected nodes) naturally emerge in the network. Each link has a specific width $w_{ij}$ and the total energy $\varepsilon_i$ associated to the site $i$ will be given by half of the sum over all link widths  connected to the site $i$ (see zoom of site $i$).}
\label{fig:sample}
\end{figure}

In this work, we introduce and study a geographically located $d$-dimensional network model. 
One of the main characteristics of this model is the possibility to control the long/short range nature of the interactions between the sites. BG statistics completely fails to describe systems that interact at long-range, and many theories have been proposed to approach this regime. 
The present model introduces a new property for this class of systems. In addition to the fact that Euclidean distances ($d_{ij}$) between the sites are relevant, it also takes into account the weights ($w_{ij}$) of the links (see Fig.~\ref{fig:sample}) and associates them to the 'energy' ($\varepsilon_i$) of each site. Due to that new ingredient, we could compute the energy distribution of the ever growing network. This distribution turns out to have the functional form of the $q$-generalised BG distribution for nonextensive systems, based on nonadditive entropy. 
These numerical results strongly suggest 
a neat correspondence between the random network geometrical problem and a particular thermostatistical problem within the generalised theory.

\section*{The model}
Our growing $d$-dimensional network starts with one site at the origin. We then stochastically locate a second site (and then a third, a fourth, and so on up to N) through a probability  $p(r) \propto 1/r^{d+ \alpha_G} \;\;\; (\alpha_G > 0)$, 
where $r\geq1$ is the Euclidean distance from the newly arrived site to the center of mass of the pre-existing cluster; $\alpha_G$ is the {\it growth} parameter and $d=1,2,3$ is the dimensionality of the system (large $\alpha_G$ yields geographically concentrated networks). 

The site $i=1$ is then linked to the site $j=2$. We sample a random number $w_{ij}$ from a distribution $P(w)$ that will give us the corresponding link weight. Each site will have a \emph{total energy} $\varepsilon_i$ that will depend on how many links it has, noted $k_i$, and the widths $\{ w_{ij}\}$ of those links. At each time step, the site $i$ only has access to its \emph{local energy} $\varepsilon_i$ defined as:

\begin{eqnarray}\label{eq:energy_site}
\varepsilon_i \equiv \sum_{j=1}^{k_i} \frac{w_{ij}}{2} \;\;\;(w_{ij} \ge0)
\end{eqnarray}
The value of $\varepsilon_i$ will directly affect the probability of the site $i$ to acquire new links. Indeed, from this step on, the sites $i =3,4, ...$ will be linked to the previous ones with probability
\begin{eqnarray}
\Pi_{ij}\propto \frac{\varepsilon_{i}}{d^{\,\alpha_A}_{ij}} \;\;(\alpha_A \ge 0)\,,
\label{attachment}
\end{eqnarray}
where $d_{ij}$ is the Euclidean distance between $i$ and $j$, where $j$ runs over all sites linked to the site $i$. The {\it attachment} parameter $\alpha_A$ controls the importance of the distance in the preferential attachment rule (\ref{attachment}). When $\alpha_A \gg 1$ the sites tends to connect to close neighbours, whereas $\alpha_A \simeq 0$ tends to generate distant connections all over the network. Notice that, while the network size increases up to $N$ nodes, the variables $k_i$ and $\varepsilon_i$ (number of links and \emph{total energy} of the $i$-th node; $i=1,2,3 \dots,N$) also increase in time (see Fig.~\ref{fig:sample} for a sample of the ever growing network).

If we consider the particular case $P(w)=\delta(w-1)$, where $\delta(z)$ denotes the Dirac delta distribution, Eq. (\ref{attachment}) becomes $\Pi_{ij}\propto k_{i}/d^{\,\alpha_A}_{ij} \;\;(\alpha_A \ge 0)\,$, 
thus recovering the usual preferential attachment rule. Consequently, the present model recovers the one in \cite{BritoSilvaTsallis2016,NunesBritoSilvaTsallis2017,BritoNunesSilvaTsallis2019} as a particular instance. Note that, if we additionally consider the particular case $\alpha_A=0$, we recover the standard Barabási-Albert model with $\Pi_i \propto k_i$~\cite{BarabasiAlbert1999, AlbertBarabasi2002}.

We are considering here the case where $w$ is given by the following stretched-exponential distribution:

\begin{eqnarray}\label{eq:w}
P(w) = \frac{\eta}{w_0\,\Gamma\left(\frac{1}{\eta}\right)} e^{-(w/w_0)^{\eta}} \;\;(w_0 >0;\, \eta > 0)\,,
\end{eqnarray}
which satisfies $\int_0^\infty dw\,P(w)=1$.
As particular cases of Eq.~(\ref{eq:w}) we have: $\eta = 1$, which corresponds to an exponential distribution, $\eta = 2$, which corresponds to a half-Gaussian distribution, and $\eta \to \infty $, which corresponds to an uniform distribution within $w \in [0,w_0] $.

\begin{figure}
\centering
\includegraphics[scale = 1]{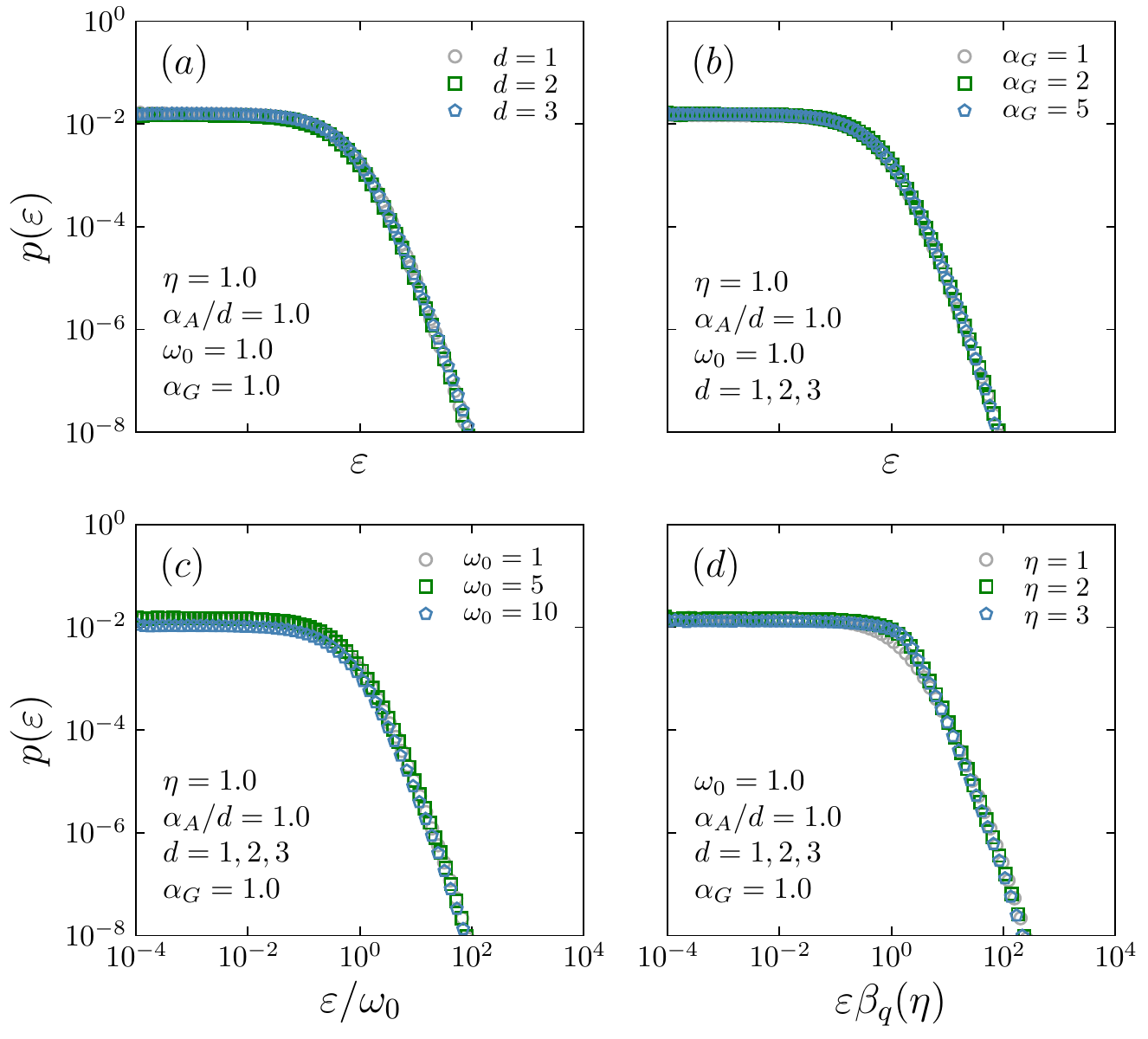}
\caption{In these plots we show $p(\varepsilon)$ for typical values of  $d$ $(a)$,  $\alpha_G$ $(b)$,  $w_0$ $(c)$ and  $\eta$ $(d)$. $(a)$ By fixing $(\alpha_G,\eta,w_0,\alpha_A/d) = (1,1,1,1)$ the dimensionality $d$ does not modify $p(\varepsilon)$. $(b)$ By fixing $(\eta,w_0,\alpha_A/d) = (1,1,1)$, $\alpha_G$ has no influence on $p(\varepsilon)$. $(c)$ We show that variations of $w_0$ yield a  $p(\varepsilon)$ which remains invariant when expressed in terms of $\varepsilon/w_0$.  $(d)$ We show that for variations of $\eta$ the curves of $p(\varepsilon)$ versus $\varepsilon\beta_q(\eta)$ collapse once again. For simplicity, the values of the fixed variables were set equal to unity, but the results remain independent from this choice. The numerical precision of all the collapses is verified to be quite impressive. Very tiny discrepancies might be due to the fact that both $N$ and the number of realisations are finite, and/or to high-order metric-topological terms. The simulations were averaged over $10^3$ realisations for $N=10^{5}$.}
\label{fig:pe_dg}
\end{figure}

\section*{Results}
Our focus here is to analyse the energy distribution $p(\varepsilon)$ of the $N \gg 1$ network. 
We have in fact analyzed a large amount of typical cases in the space $(d, \alpha_A, \alpha_G, w_0,\eta)$, and have systematically found the same results for $d=1,2,3$ within the intervals $(\alpha_A/d \in [0,10]; \alpha_G \in [1,10]; w_0 \in [0.5, 10]; \eta \in [0.5,3])$.
Similarly to previous works \cite{SoaresTsallisMarizSilva2005, BritoSilvaTsallis2016, NunesBritoSilvaTsallis2017, BritoNunesSilvaTsallis2019}, 
$p(\varepsilon)$ does not depend on $\alpha_G$; also, it does not depend independently on $d$ and $\alpha_A$, but only, remarkably, on the ratio $\alpha_A/d$;  $p(\varepsilon)$ also depends on $w_0$ and $\eta$ (see Fig.~\ref{fig:pe_dg}$(a)$-$(d)$).
Because of these features, and without loss of generality, we have once for ever fixed $\alpha_G = 1$, and $d=2$. The simulations were done for $10^3$ realisations of size $N=10^{5}$, which was verified to be enough for observing the asymptotic distribution $p(\varepsilon)$ with high precision. 

\begin{figure}[!h]
\centering
\includegraphics[scale=1.2]{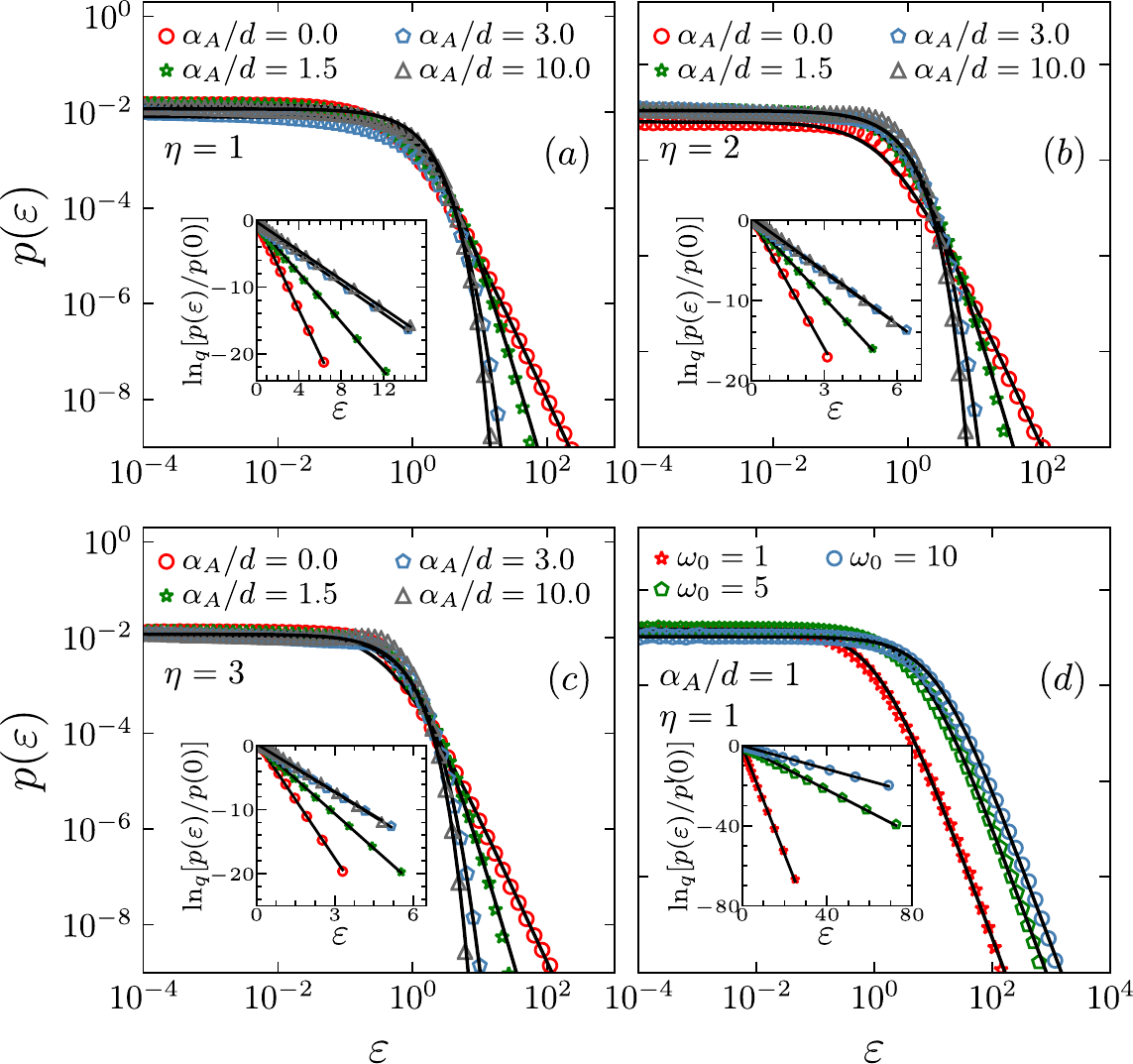}
\caption{In these plots we show the variations of $p(\varepsilon)$ for fixed values $\eta = 1$ $(a)$ ,  $\eta = 2$ $(b)$ and  $\eta = 3$ $(c)$, for $\alpha_A/d = 0,1.5, 3,10$. In  $(d)$ we show the variations of $p(\varepsilon)$ for fixed values of $(\alpha_A/d,\eta) = (1,1)$ and $w_0 = 1,5,10$.  In all figures, the black continuous lines are given by Eq.~(\ref{eq:pe}) with $(q,\beta_q)$ given by Eqs.~(\ref{eq:q}, \,\ref{eq:beta}) respectively. {\it Insets:} $\ln_q$-linear representation of the same data; the slopes of the straight lines precisely yield the corresponding values of ($-\beta_q$). The simulations were averaged over $10^3$ realisations for $N=10^{5}$.}
\label{fig:pe}
\end{figure}

We know that the signature of the Boltzmannian systems is the presence of exponentials and Gaussians distributions. Similarly, the nonextensive systems based on the entropy $S_q$ can be recognised by the emergence of $q$-exponentials and $q$-Gaussians distributions. We have here found that, independent of the choice of $(d, \alpha_A, \alpha_G, w_0,\eta)$, the \emph{'energy' distribution} $p(\varepsilon)$ associated with the network is {\it invariably} well fitted  by the following $q$-exponential:
\begin{eqnarray}
p_q(\varepsilon) =\frac{e_q^{-\beta_{q} \varepsilon}}{Z_q}\label{eq:pe},
\end{eqnarray}
where $p_q(\varepsilon)$ represents the generalisation, within nonextensive statistical mechanics, of the BG energy weight  with $\varepsilon$, $\beta_q$ and $Z_q$ playing respectively the roles of energy, inverse temperature and normalisation factor (see Fig.~\ref{fig:pe}). Note that, when $q\to 1$, we do recover the standard Boltzmann distribution since  $e_1^{-\beta_1 \varepsilon}\equiv e^{-\beta \varepsilon}$. This result exhibits an interesting emergence of 
correspondence between a random network geometric problem and a particular case within generalised thermostatistics. This fact definitively reminds the Kasteleyn-Fortuin theorem \cite{KasteleynFortuin1969}, which establishes an important isomorphism between the bond-percolation geometric problem and the $q_{Potts} \to 1$ limit of the $q_{Potts}$-state Potts ferromagnet.

For all $(d, \alpha_A, \alpha_G, w_0, \eta)$, we found that:
\begin{eqnarray}
q =
\begin{cases}
\frac{4}{3} &\mbox{if}\; 0\leq \frac{\alpha_A}{d} \leq 1\\
\frac{1}{3}\,e^{1-\alpha_A/d} + 1 &\mbox{if}\; \frac{\alpha_A}{d} > 1
\end{cases}
\label{eq:q}
\end{eqnarray}

\begin{eqnarray}
\beta_q =
\begin{cases}
\displaystyle\beta_{q_0} &\mbox{if } \leq \frac{\alpha_A}{d} \leq 1\\
\displaystyle (\beta_{q_0} - \beta_{q_\infty})\,e^{2(1-\alpha_A/d)} +\beta_{q_\infty}&\mbox{if } \frac{\alpha_A}{d} > 1
\end{cases}
\label{eq:beta}
\end{eqnarray}
with $\beta_{q_0} \simeq (-10.81 e^{-1.36\eta} + 6.04)/w_0$ and $\beta_{q_\infty} \simeq (-4.81 e^{-1.22\eta} + 2.56)/w_0$ . As can be seen, $q$ does not depend on $(\eta,w_0)$, but only on the scaled variable $\alpha_A/d$. In contrast, $\beta_q$ is less universal and depends on all three parameters $(w_0,\eta,\alpha_A/d)$.

\begin{figure}[!h]
\centering
\includegraphics[width=\textwidth]{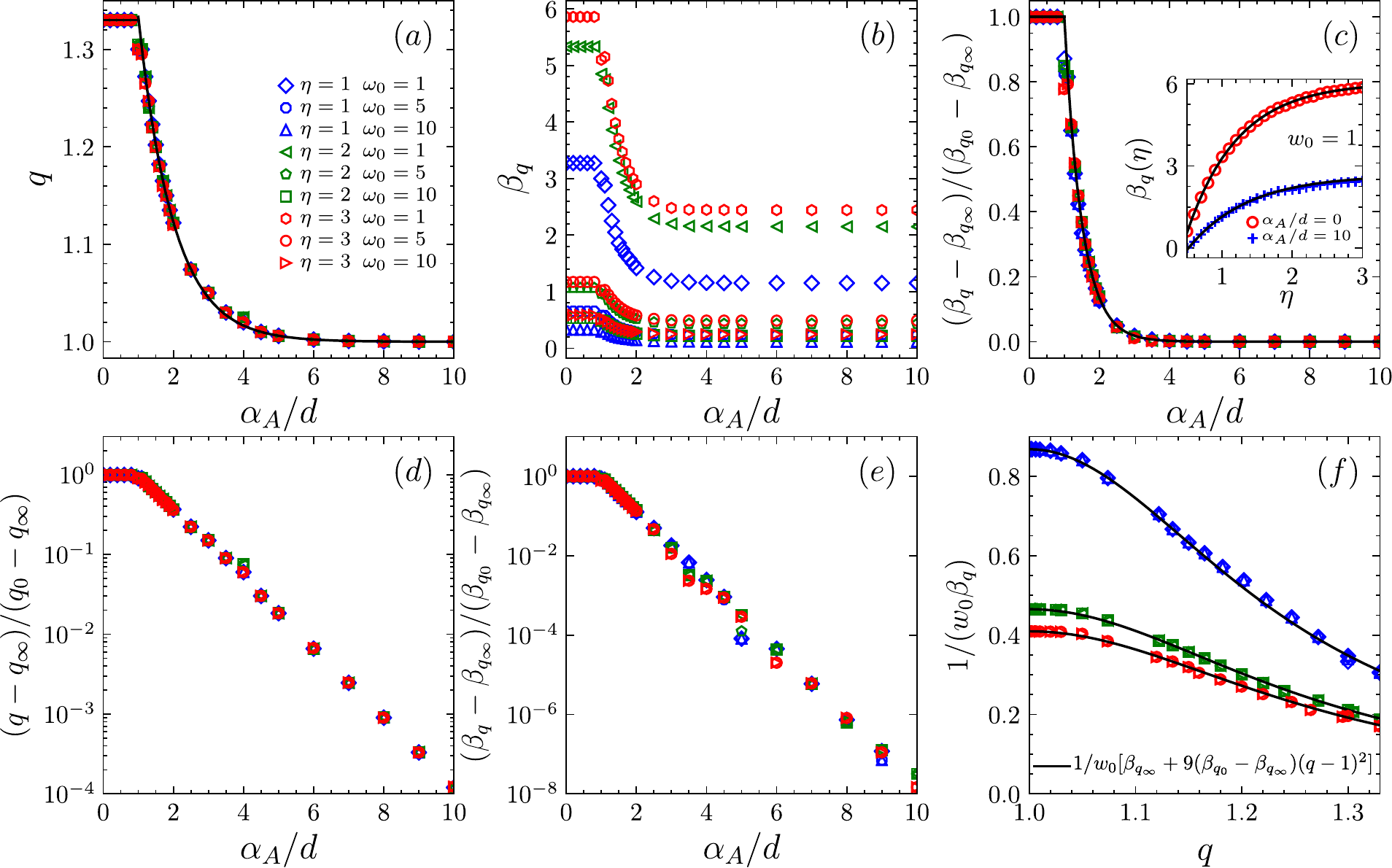}
\caption{$(a)$ $q$ as a function of $\alpha_A/d$; $q$ is constant in the range $0\leq \alpha_A/d \leq 1$ ($q_0 = 4/3$) and  decreases exponentially with $\alpha_A/d$ for $\alpha_A/d > 1$, down to $q_\infty = 1$ (black solid line). $(b)$ $\beta_q$ as a function of $\alpha_A/d$ for $\eta = 1, 2, 3$ and $w_0 = 1, 5,10$, for typical values of $\alpha_A/d$; $\beta_q$ increases with $\eta$ and decreases with $w_0$ and $\alpha_A/d$. $(c)$ By plotting  $(\beta_{q} - \beta_{q_\infty})/(\beta_{q_0} - \beta_{q_\infty})$, all curves collapse and exponentially decrease with $\alpha_A/d >1$ (black straight line). {\it Inset:} $\beta_{q_0}$ and $\beta_{q_\infty}$ exponentially vary with $\eta$;  $\beta_{q_\infty}$ was estimated by fixing $\alpha_A/d =10$. 
In $(d)$ and $(e)$ we present $\log$-linear representations of the same data as in $(a)$ and $(c)$ 
respectively, exhibiting the exponential dependence of both $q$ and $\beta_q$ on $\alpha_A/d$, when $\alpha_A/d \geq 1$. 
$(f)$ By eliminating the variable $\alpha_A/d$, we show $1/(w_0\beta_q)$ as functions of $q$ for the same set of data shown in the previous plots; $q$ is related with $1/(w_0\beta_q)$ through the equation $1/w_0[\beta_{q_\infty} + 9(\beta_{q_0} - \beta_{q_\infty})(q-1)^2]$ that is valid for all values of $(w_0,\eta$). For very large values of $\alpha_A/d$ and the extreme regions $\eta \to 0$ and $\eta \to \infty$, the  numerical precision needed to attain the stationary-state distribution is too high for our present computational effort and further analysis would be needed, which is out of the present scope.
}
\label{fig:q-b}
\label{fig:q-b}
\end{figure}


In Fig.~\ref{fig:q-b}(a,d) we show the numerical results for the index $q$ as function of $\alpha_A/d$. This result is consistent with \cite{BritoSilvaTsallis2016, BritoNunesSilvaTsallis2019}, where the behaviour of $q$ characterises the existence of three regimes. As can be seen, $q$ is constant and equal to $4/3$ in the range $0 \leq \alpha_A/d \leq 1$. This interval describes the regime of strong-long-range interactions characterised by the highest value of $q$. In the interval $1< \alpha_A/d \lesssim 5$ we have the moderately-long-range interaction regime, characterised by $q$ smaller than $4/3$ but still greater than $1$. In this regime $q$  displays no abrupt transition from $4/3$ to $1$ but instead it decreases exponentially with $\alpha_A/d$ through the function $e^{1-\alpha_A/d}$ \cite{BritoSilvaTsallis2016, BritoNunesSilvaTsallis2019}. This behaviour exhibits that the BG regime was not yet reached. In the last regime, $\alpha_A/d \gtrsim 5$, the Boltzmannian-like regime finally emerges and $q = 1$. Similar results for $q$ were found in \cite{ThurnerTsallis2005} for a gas-like network model. 
In the Fig.~\ref{fig:q-b} (b, c,e) we show similar results for the parameter $\beta_q$ which equals $\beta_{q_0}$ in the range $0\leq\alpha_A/d\leq 1$ and, then, exponentially decreases with $\alpha_A/d$; $\beta_q$ increases with $\eta$ and decreases with $w_0$. However, if we plot $ (\beta_{q} - \beta_{q_\infty})/(\beta_{q_0} - \beta_{q_\infty})$, all curves collapse as a function of $\alpha_A/d$. Moreover, we verify in Fig. \ref{fig:q-b}(f) that the effective 'temperature' {\it decreases} with increasing $q$. There is no thermodynamical prescription which would impose that. In \cite{GrecoTsallisRapisardaPluchinoFicheraContrafatto2020}, for instance, both possibilities are in fact observed.

\section*{Discussion}

All these results strongly suggest that the 'energy' distribution $p(\varepsilon)$ of the network is given by the very same expression which $q$-generalises the Boltzmann-Gibbs weight when it is the nonadditive entropy $S_q$ which is optimised. Naturally, since the present study is numerical, we can not exclude very minor corrections due to high-order metric-topological terms. 
The BG limit is rapidly reached when $\alpha_A/d\gtrsim 5$.  Not less important, $q$ and $\beta_q$ depend on $\alpha_A$ and $d$ only through the ratio $\alpha_A/d$; also, interestingly enough, none of them depends on $\alpha_G$. The fact that $q$ depends {\it only} on $\alpha_A/d$ means that this ratio uniquely determines the entropic nonadditivity universality class. The quantity $\beta_q$ also depends on $\eta$ and $w_0$. Consistently, $w_0$, which characterises the width of the stretched-exponential distribution $P(w)$, plays here the same role as $T$ in usual thermal BG problems. This seemingly is the first time that, in a complex network, we identify a parameter which plays the role of an external parameter that we may fix at will, similarly to the manner in which we fix, in BG statistical mechanics, the temperature at which the thermally equilibrated system is placed. In all previous connections with random networks \cite{SoaresTsallisMarizSilva2005,ThurnerTsallis2005,Thurner2005,BritoSilvaTsallis2016,NunesBritoSilvaTsallis2017,BritoNunesSilvaTsallis2019,CinardiRapisardaTsallis2020}, $\beta_q$ (sometimes noted $1/\kappa_q$) was univocally related to $q$. A single value for $\beta_q$ for a given value of $q$ is analogous to traditional critical points in BG statistics. In our present case, we have, for a fixed value of $q$, the freedom of making $\beta_q$ to vary, like $T$ in BG thermal statistics.

The present results strongly support the conjecture of existence of 
a neat correspondence between geometrical random network (asymptotically) scale-invariant problems and the present specific class of many-body models within nonextensive statistical mechanics, constructed upon nonadditive entropies. This is analogous to the Kasteleyn-Fortuin theorem for the $q_{Potts}$-state Potts model, whose $q_{Potts}\to 1$ limit rigorously corresponds to the bond percolation problem~\cite{KasteleynFortuin1969}, and also to the de Gennes celebrated isomorphism for the $n$-vector ferromagnetic model, whose  $n\to 0$ limit precisely corresponds to the  self-avoiding random walk~\cite{deGennes1972}, which constitutes a pillar in polymer physics. It is possible to think of a variety of applications of connections of the present kind, for example the maintaining budgets to be distributed among cities connected within a large regional network of roads. Each city could, for instance, receive a support proportional to the sum of the widths of the roads arriving to it.   

\section*{Acknowledgements}

S. B. acknowledges the Serrapilheira Institute (Grant No. Serra-1708-15763), the Brazilian agencies MCTI and MEC. R.M.O., L.R.S. and C.T. acknowledge partial financial support from CAPES, CNPq and Faperj (Brazilian agencies). 
We also thank the High Performance Computing Center (NPAD/UFRN) for providing computational resources.

\section*{Methods}
To calculate the relevant properties of our network model in a statistically relevant way, we used $1000$ independent realisations within the standard Monte Carlo method to generate different instances of the our network.  The network size was set to be $N=10^5$. All simulations were obtained through independent codes in C. To generate random numbers from the stretched exponential distribution we used the boost library available in \url{https://www.boost.org/}. Logarithmic binning was used to generate the histogram of the energies using Python packages of the \emph{numpy} library



\end{document}